\documentclass[10pt,twoside]{hsqcd}
\usepackage{epsf,amsmath}
\usepackage{graphicx}

\newcommand{\bce}{\begin{center}}
\newcommand{\ece}{\end{center}}
\newcommand{\be}{\begin{equation}}
\newcommand{\ee}{\end{equation}}
\newcommand{\bea}{\begin{eqnarray}}

\newcommand{\eea}{\end{eqnarray}}

\newcommand{\ba}{\begin{array}}
\newcommand{\ea}{\end{array}}

\newcommand{\brho}{\mbox{\boldmath $\rho$}}

\newcommand{\bDelta}{\mbox{\boldmath $\Delta$}}

\newcommand{\bkappa}{\mbox{\boldmath $\kappa$}}

\newcommand{\bb}{{\bf b}}
\newcommand{\br}{{\bf r}}

\newcommand{\bk}{{\bf k}}

\newcommand{\bp}{{\bf p}}
\newcommand{\bq}{{\bf q}}

\def\lsim{\mathrel{\rlap{\lower4pt\hbox{\hskip1pt$\sim$}}
    \raise1pt\hbox{$<$}}}         
\def\gsim{\mathrel{\rlap{\lower4pt\hbox{\hskip1pt$\sim$}}
    \raise1pt\hbox{$>$}}}         
\def\beq{\begin{equation}}
\def\eeq{\end{equation}}
\def\bea{\begin{eqnarray}}
\def\eea{\end{eqnarray}}

\def\lsim{\mathrel{\rlap{\lower4pt\hbox{\hskip1pt$\sim$}}
    \raise1pt\hbox{$<$}}}         

\def\gsim{\mathrel{\rlap{\lower4pt\hbox{\hskip1pt$\sim$}}
    \raise1pt\hbox{$>$}}}         

\begin{document}

\title{HARD SCATTERING IN A NUCLEAR ENVIRONMENT: FAREWELL TO LINEAR $k_{\perp}$-FACTORIZATION
\thanks{This work was
supported in part by the INTAS grant 00-0036 }}

\author{\underline{  N.N. Nikolaev}$^{a,b)}$,
W. Sch\"afer$^{a)}$, B.G. Zakharov$^{b)}$, V.R.
Zoller$^{d)}$\medskip\\  
$^{a)}$ Institut f. Kernphysik, Forschungszentrum J\"ulich, D-52425 J\"ulich, Germany\\
$^{b)}$ L.D.Landau Institute for Theoretical Physics, 142432 Chernogolovka, Russia\\
$^{d)}$ Institute for Theoretical and Experimental Physics, 117259 Moscow, Russia\\
E-mail: N.Nikolaev@fz-juelich.de  }

\maketitle

\begin{abstract}
\noindent We discuss a dramatic change brought into the pQCD description of 
hard processes in a nuclear environment by a large thickness of heavy nuclei. 
It breaks the familiar linear $k_{\perp}$-factorization which must
be replaced by a new concept of the nonlinear $k_{\perp}$-factorization
introduced in \cite{Nonlinear}.We demonstrate the salient features
of nonlinear $k_{\perp}$-factorization on several examples from hard dijet
production in DIS off heavy nuclei to single-jet to dijet production
in hadron-nucleus collisions . We also comment briefly on the 
non-linear BFKL evolution for gluon density of nuclei.
\end{abstract}



\markboth{\large \sl N.~N.~Nikolaev et al. \hspace*{2cm} HSQCD 2004}
{\large \sl \hspace*{1cm}  HSQCD'04 PROCEEDINGS}


\section{Introduction}
The linear $k_{\perp}$-factorization is a fundamental ingredient of 
the pQCD description of high energy hard processes off free nucleons.
A large thickness 
of a target nucleus introduces a new scale - the so-called saturation 
scale $Q_A^2$, - which breaks the linear $k_{\perp}$-factorization 
theorems for hard scattering in a nuclear environment. This property 
can be linked to the
unitarity constraints for the color dipole-nucleus interaction. 
In this talk we review the recent work by the ITEP-J\"ulich-Landau
collaboration in which a new concept of the nonlinear 
$k_{\perp}$-factorization has been introduced 
\cite{Nonlinear,PionDijet}. We illustrate this new concept on an 
example of dijet production in DIS off heavy nuclei and comment on more 
recent applications to single-jet \cite{SingleJet} and dijet 
\cite{pAdijet} production in $pA$
collisions. The nonlinear $k_{\perp}$-factorization emerges as a
generic feature of the pQCD approach to hard processes in nuclear 
environment, although the concrete realizations depend strongly
on the relevant pQCD subprocesses.

\section{The $k_{\perp}$-factorization for DIS off free nucleons}
The parton fusion  
description of the shadowing introduced in 1975 \cite{NZfusion} is 
equivalent to the unitarization on the color dipole-nucleus interaction 
\cite{NZ91}.
The starting point is the color dipole 
factorization for DIS at small $x \lsim x_A=1/R_A m_N$, when 
the coherency over the thickness of
the nucleus holds for the $q\bar{q}$ Fock states of the virtual photon: 

\bea
\sigma_T(x,Q^2) = 
\langle {{\gamma^*}}|
{{\sigma(x,\br)}} |{{\gamma^*}} 
\rangle 
 =\int_{0}^1 dz \int {{d^2\br}}
{{\Psi^*_{{{\gamma^*}}}(z,{{\br}})}}
 {{\sigma(x,\br)}} {
{\Psi_{{{\gamma^*}}}(z,{{\br}})}}\,.
\label{eq:2.1}
\eea
Here $z$ and
$(1-z) $ is the energy partition between {{$q$}}
 \& ${{\bar{q}}}$ and ${{\br}}=$ size of the color dipole.
There is an  {{equivalence}} between
{{color dipole}} and
{{$k_{\perp}$-factorization}} \cite{NZ91,NZglue,NZ94}:
\bea
{{\sigma(x,\br)}}&=& \alpha_S(r)
\int \frac{d^2{{\bkappa}}4\pi
[1-e^{i{{\bkappa}} \br }] }{
N_c{{\kappa}}^4} \cdot {{\frac{\partial
G_N}
{\partial\log{{\kappa}}^2} }} \,,
\label{eq:2.2}\\
 f(x,{{\bkappa}} ) &=& \frac{4\pi}
{N_c\sigma_0(x)}\cdot \frac{1}
{{{\kappa}}^4} \, \cdot
{{\frac{\partial G_N(x,{{\bkappa}})}
{\partial\log{{\kappa}}^2} }}\, .\label{eq:2.3}
\eea
where $ \sigma_0(x) = {{\sigma(x,\br)}}\big|_{{{r\to \infty}}}$. The
$x$-dependence of ${{\sigma(x,\br)}}$ is governed by the 
color dipole BFKL equation \cite{NZZJETPLett}.
The unintegrated gluon density 
${{f(x,\bkappa)}}$ furnishes a {{universal}} description of 
$F_{2p}(x,Q^2)$ and of the final states. For instance,
the linear {{$k_{\perp}$-factorization}} for forward dijet
cross section reads
\bea
\frac{(2\pi)^2d\sigma_N}
{dz {{d^2\bp_+}}
{{ d^2\bDelta}}} = 
 \frac{ \alpha_S({{\bp^2}}) }
{2}
 f(x,{{ \bDelta}} )
\times 
\left|\Psi(z,\bp_+) -
\Psi(z,{{\bp_+}} -
{{\bDelta }})\right|^2 \,,
\label{eq:2.4}
\eea
where $\Psi(z,\bp)$ is the $q\bar{q}$ wave function of the photon
and 
${{\bDelta}}={{\bp_+}} +
{{\bp_-}}$ is the jet-jet decorrelation momentum. 


\section{Collective unintegrated nuclear glue}

The color dipole-nucleus cross-section 
\cite{NZ91} 
$
{{\sigma_A(\br)}} = 2\!\!\int\!\!
d^2{{\bb} }[1 -\exp(-\frac{1}
{2}
{{\sigma(\br)}}
T({{\bb}}))]$\,
where $T(\bb)$ is the optical the
thickness of a nucleus, 
defines the collective nuclear glue per
{{ unit area}} in the impact parameter space,
${{\phi}}({{\bb}},x,{{\bkappa}}
) $
\cite{NSSdijet,Nonlinear}:

\begin{figure}[!thb]
\vspace*{3.0cm}
\begin{center}
\includegraphics{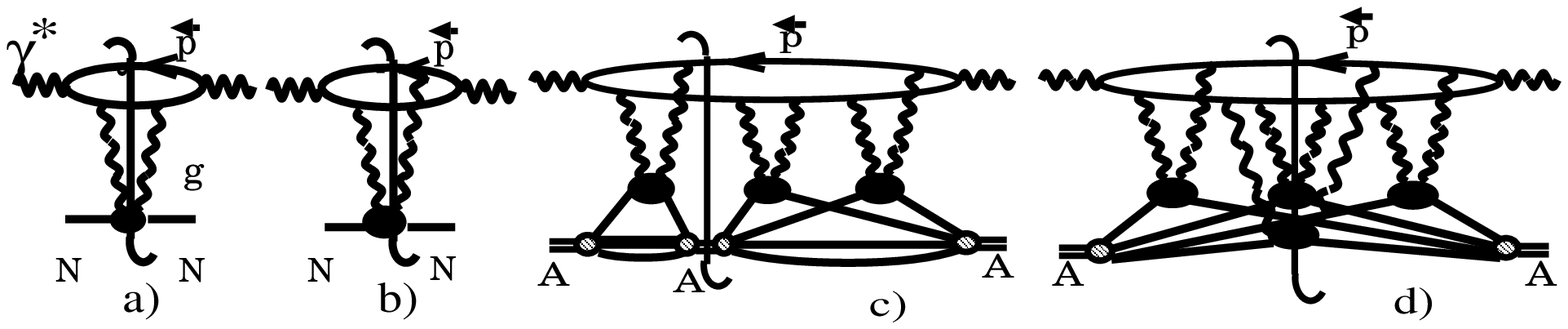}
\caption[*]{ The typical unitarity cuts and dijet final states in DIS : (a),(b) - 
free-nucleon target, (c) - coherent diffractive DIS off a nucleus,
(d) - truly inelastic DIS with multiple color excitation of the nucleus.}
\label{fig:UnitarityDiffraction2004}
\end{center}
\end{figure}

\bea \Gamma_A({{\bb}},{{\br}})&=&[1 -\exp(-\frac{1}
{2}
{{\sigma(\br)}}T({{\bb}}))]
=\int d^2{{\bkappa}}
{{\phi}}({{\bb}},x,{{\bkappa}}
) \{1-\exp[i {{\bkappa}}
{{\br}} ]\} \,. \label{eq:3.1} \eea
A useful expansion is
\bea
{{\phi}}({{\bb}},x,{{\bkappa}}
) = \sum_{j=1}^{\infty}{{
w_{j}({{\bb}})}}
 f^{(j)}(x,{{\bkappa}} )\, ,
~~~{{w_{j}({{\bb}})}} =\frac{
1}
{j!} \left[\frac{1}
{2}T({{\bb}})\right]^j \exp\left[-\nu_{A}(x,{{\bb}})\right],
\label{eq:3.2} \eea
where $\nu_{A}(x,{{\bb}}) =  
\frac{1}
{2}\sigma_0(x) {{T(\bb)}}$,
 ${{w_{j}}}$ is the probability
to find {{ $j$ overlapping nucleons}} at impact
parameter ${{\bb}}$ in a Lorentz-contracted
nucleus and  $f^{(j)}(x,\bkappa)$ is a {{collective glue of $j$
overlapping nucleons}}:
\bea f^{(j)}(x,{{\bkappa}} )\,=\,  \int
\prod_{i=1}^j d^2{{\bkappa}}_{i}
f(x,{{\bkappa}} _{i})
\delta({{\bkappa}} -\sum_{i=1}^j
{{\bkappa}} _i)\,~~~ f^{(0)}(x,{{\bkappa}} )= \delta({{\bkappa}} )
\label{eq:3.3} \eea
The plateau at small momenta of gluons, 
\bea
{{\phi}}({{\bb}},x,{{\bkappa}})
\approx \frac{1}
{\pi}
\frac{ Q_A^2(\bb)} 
{({{\bkappa}}^2+ {{
Q_A^2}}({{\bb}}))^2}\, , 
~~~{{ Q_A^2}}({{\bb}},x) \approx
 \frac{4\pi^2}
{N_c} \alpha_S({{
Q_A^2}})G(x,{{ Q_A^2}}) T({{\bb}})\,,
\label{eq:3.4} \eea
is a signal of the saturation effect.
The collective nuclear glue furnishes the linear
$k_{\perp}$-factorization representation  
for DIS off nuclei (hereafter we focus on $x\lsim x_A$),
\bea
\sigma_{{{\gamma^*}}A}& =& 
\int {{d^2\bb}}\langle {{\gamma^*}} | 2\{1- \exp[-\frac{1}{2}{{\sigma({{\br}})}}{{T(\bb)}}]\}| 
{{\gamma^*}}\rangle 
\nonumber\\
&=&\int {{d^2\bb}} \int \frac{d^2{{\bp}}} {(2\pi)^2} \alpha_S({{\bp}}^2)
\int  d^2{{\bkappa}}  {{{{\phi}}}}(\bb,x_A,{{\bkappa}} )
(\Psi(z,{{\bp}}) - \Psi(z,{{\bp}}-{{\bkappa}} ))^2
\label{eq:3.5}
\eea
which is the same as for the free-nucleon target, subject to
$f(x_A,{{\bkappa}}) {{\Longleftrightarrow}}
 {{{{\phi}}}}(\bb,x_A,{{\bkappa}})$.

\begin{figure}[!thb]
\vspace*{3.2cm}
\begin{center}
\includegraphics{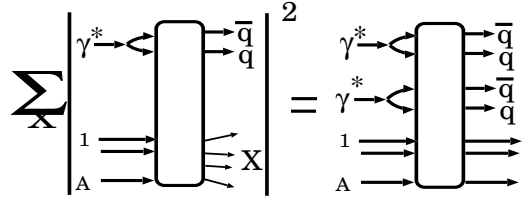}
\caption[*]{Unitarity diagram for the dijet spectrum in terms
of the 4-parton scattering amplitude. }
\end{center}
\label{fig:4BodyUnitarity}
\end{figure}


\section{The non-abelian intranuclear evolution of color dipoles}
 
The two typical final states in DIS off heavy nucleus are shown
in fig.~\ref{fig:UnitarityDiffraction2004}. The coherent diffraction
with large rapidity gap between the target nucleus in the ground state
and diffractive hadronic 
debris of the photon makes $\approx 50\%$ of the total cross section
\cite{NZZdiffr}
and gives exactly back-to-back correlated dijets. In the 
truly inelastic DIS with multiple color excitation of the nucleus
one encounters the non-Abelian intranuclear evolution of color 
dipoles, the consistent description of which 
based on the ideas from 
\cite{SlavaPositronium,NPZcharm} is found in \cite{Nonlinear} .
\noindent
Specifically, the {\it ab 
initio} calculation of the nuclear distortion of the
two-parton density matrix the Fourier transform of which gives the
spectrum of dijets, can be reduced, upon the closure over nuclear excitations, 
to the problem of intranuclear
propagation of the color-singlet 4-parton states as illustrated
in fig. \ref{fig:4BodyUnitarity}: 

\bea
\frac{(2\pi)^4d\sigma_{in}} { dz d^2{{\bp_+}} d^2{{\bp_-}}}& =&
 \int d^2 {{\bb_+}}' d^2{{\bb_-}}' 
d^2{{\bb_+}} d^2{{\bb_-}}  \exp[-i{{\bp_+}}({{\bb_+}} -{{\bb_+}}')-i{{\bp_-}}({{\bb_-}} -
{{\bb_-}}')]\nonumber\\
&\times &\Psi^*(Q^2,z,{{\bb_+}}' -{{\bb_-}}')
\Psi(Q^2,z,{{\bb_+}} -{{\bb_-}})\nonumber\\
&\times& \Bigl\{S_{4A}({{\bb_+}}',{{\bb_-}}',{{\bb_+}},{{\bb_-}})- S_{4A}^{(Diffr)}({{\bb_+}}',
{{\bb_-}}',{{\bb_+}},{{\bb_-}}) \Bigr\}\, ,
\label{eq:4.1}
\eea
where we subtracted the diffractive contribution. To
the standard dilute-gas nucleus approximation, the
Glauber-Gribov theory gives
 \bea S_{4A}({{\bb_+}}',{{\bb_-}}',{{\bb_+}},{{\bb_-}})=\exp\{-
\frac{1}{2}\sigma_{4}({{\bb_+}}',{{\bb_-}}',{{\bb_+}}',{{\bb_-}})
T({{\bb}})\}\, .
\label{eq:4.2} 
\eea 
where $\sigma_{4}({{\bb_+}}',{{\bb_-}}',{{\bb_+}}',{{\bb_-}})$ is the coupled-channel operator
in the space of {{singlet-singlet}} $|{{11}}\rangle$
 or  {{ octet-octet}} $|{{88}}\rangle$
4-body dipoles, see
ref. \cite{Nonlinear} for more details.


\section{The fate of $k_{\perp}$-factorization for nuclear targets: the case of DIS }

Taken separately, both the truly inelastic, 
\bea
\frac{(2\pi)^2d \sigma_{in}}{d^2\bb d^2{{\bp}} dz }  & =& 
\int  d^2{{\bkappa}} 
{{\phi(\bb,x_A,{{\bkappa}} )}}
\bigl|\Psi(z,{{\bp}}) -
\Psi(z,{{\bp}}-
{{\bkappa}}) \bigr|^2  \nonumber\\
&-&
 {{\Bigl|}}
\int d^2{{\bkappa}}
{{\phi(\bb,x_A,{{\bkappa}})}} 
(\Psi(z,{{\bp}}) -
\Psi(z,{{\bp}}-
{{\bkappa}})){{\Bigr|^2}}
\label{eq:5.1}
\eea
and coherent diffractive 
\bea
\frac{(2\pi)^2d \sigma_{D}}{ d^2\bb d^2{{\bp}} dz }   =
\left|\int d^2{{\bkappa}}{
{\phi(\bb,x_A,{{\bkappa}})}} 
(\Psi(z,{{\bp}}) -
\Psi(z,{{\bp}}-{{\bkappa}}))\right|^2 \, .
\label{eq:5.2}
\eea
spectra are nonlinear functionals of nuclear glue, but they add to precisely 
the differential form of eq.~
(\ref{eq:3.5}). I.e., the linear $k_{\perp}$-factorization in terms of 
the collective  nuclear glue holds for total DIS as if there were no Initial and Final State
distortions of the spectrum of leading quarks. Such and abelianization is a
feature of DIS where the photon is a color
singlet projectile, the same is not true  
for other projectiles, see below Sect. 7.
\begin{figure}[!thb]
\vspace*{4.8cm}
\begin{center}
\includegraphics{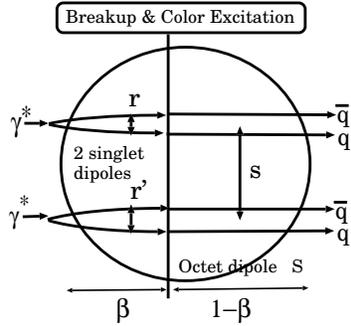}
\caption[*]{The color excitation of the dipole in the
large-$N_c$ approximation.}
\end{center}
\label{fig:Distortions}
\end{figure}

The nuclear dijet spectrum is a  manifestly {{nonlinear}} functional of the
collective nuclear glue and here emerges a concept of the
nonlinear$k_{\perp}$-
{{factorization}}(here we cite the result for the large-$N_c$ approximation):
\bea
&&\frac{(2\pi)^2d\sigma_{in} }{ d^2{{\bb}} dz d\bp_{-} d{{\bDelta}}} = \frac{1}{
2} T(\bb) 
\int_0^1 d {{\beta}}
\int d^2{{\bkappa_1}} d^2{{\bkappa}} f(x_A,{{\bkappa}})
{{\Phi}}({{(1-\beta)}}\nu_A({{\bb}}),x_A,{{\bDelta}} -{{\bkappa_1}} -{{\bkappa}})
\nonumber\\
&\times&{{\Phi}}({{(1-\beta)}}\nu_A({{\bb}}),x_A,{{\bkappa_1}})
{{\Bigl|}}
\Psi(\beta;z,\bp_{-} +{{\bkappa_1}}) -
\Psi(\beta; z,\bp_{-}  +{{\bkappa_1}}+{{\bkappa}})
{{\Bigr|^2}}
\, .
\label{eq:5.3}
\eea
where ${{\Phi}}(\nu_A(\bb),x_A,{{\bkappa}})= \delta(\bkappa)\exp(-\nu_A(\bb)) +
\phi(\bb,x_A,\bkappa)$ and 
\bea\Psi(\beta;z,\bp)= \int d^2{{\bkappa}}{{\Phi}}({{\beta}}\nu_A({{\bb}}),x_A,{{\bkappa}})
\Psi(\beta;z,\bp_{-} +{{\bkappa}})
\label{eq:5.4}
\eea
 is the wave function of the incident color-singlet
dipole distorted by the Initial State Interaction in the slice ${{\beta}}$ of
a nucleus (see fig.~\ref{fig:Distortions}).
The slice 
{{
${{(1-\beta)}}$}} in which
the dipole is in the color-octet state 
gives the Final State Interaction. In DIS it looks as an 
independent broadening of the quark
and antiquark jets. 
Evidently, eq.~(\ref{eq:5.3} entails nuclear enhancement of the decorrelation 
of jets, the {{semihard}} dijets,
$|\bp_\pm|^2 \lsim Q_A^2$, are completely decorrelated.

\section{The fate of $k_{\perp}$-factorization for nuclear targets: dijets from $pA$ collisions}

Here we comment on the recent generalization of nonlinear $k_{\perp}$-factorization
from DIS to hard interactions of hadrons and nuclei \cite{pAdijet}.
The pQCD subprocesses relevant to dijet production in $pp$ 
collisions and the proton hemisphere of $pA$ collisions at RHIC
\cite{STAR,BRAHMS} are 
collisions of a beam quark $q^*$ or a gluon $g^*$ with a gluon from the
target, $q^*g \to qg$ or $g^*g \to gg$, respectively. In contrast to
the colorless photon in DIS the fragmenting parton is a colored one.
We describe the principal changes caused by that on an example of 
fragmentation $Q^*\to qg$. The total spectrum of dijets including the
diffractive component in which the target nucleus remains in the 
ground state, will be described by eq.~(\ref{eq:4.1}) in which the last
line will be replaced by the combination of multiparton $S$-matrices
shown in fig.~\ref{fig:DensityMatrixHSQCD}. 
The intranuclear propagation of
the 2-parton and 3-parton states is a single-channel problem \cite{NPZcharm,Nonlinear},
the non-Abelian evolution of the 4-parton state is a three-channel problem,
the detailed solution of which will be published elsewhere \cite{pAdijet}. Here we
only cite the nuclear dijet spectrum found in the large-$N_c$ approximation:
\bea
\frac{(2\pi)^2d\sigma_{in} }{ d^2\bb dz d\bp_{g} d{\bDelta}} &=& \frac{1}{
2} T(\bb) 
\int_0^1 d {{\beta}}
\int d^2{{\bkappa_1}} d^2{{\bkappa}} f(x_A,{{\bkappa}})\nonumber\\
&\times&{{\Phi}}({{(2-\beta)}}\nu_A({{\bb}}),x_A{{\bDelta}} -{{\bkappa_1}} -{{\bkappa}})
{{\Phi}}({{(1-\beta)}}\nu_A({{\bb}}),x_A,{{\bkappa_1}})\nonumber\\
&\times&{{\Bigl|}}
\Psi(\beta;z_g,\bp_{g} +{{\bkappa_1}}) -
- 
\Psi(\beta; z_g,\bp_{g}  +{{\bkappa_1}}+{{\bkappa}})
{{\Bigr|^2}}\nonumber\\
&+&\phi(\nu_A(\bb),x_A,\bDelta) {{\Bigl|}}
\Psi(1;z_g,\bp_{g}) -
\Psi(z_g,\bp_{g}  +z\bDelta)
{{\Bigr|^2}}\nonumber\\
&+&\delta(\bDelta){{\Bigl|}}
\Psi(1;z_g,\bp_{g}) -
\Psi( z_g,\bp_{g})
{{\Bigr|^2}}\exp\left[-\nu_A(\bb)\right]\, ,
\label{eq:6.1}
\eea 
\begin{figure}[!thb]
\vspace*{2.8cm}
\begin{center}
\includegraphics{DensityMatrixHSQCDbw.eps}
\caption[*]{ The $\textsf{S}$-matrix structure of the two-body density
matrix for excitation $a\to bc$.}
\label{fig:DensityMatrixHSQCD}
\end{center}
\end{figure}
where $\Psi( z_g,\bp_{g})$ is the momentum-space WF of the $qg$ Fock state
of the quark $Q^*$. This must be compared to the linear $k_{\perp}$-factorization
for the free-nucleon target at large $N_c$: 
\bea
&&\frac{(2\pi)^2d\sigma_N }{ dz {{d^2\bp_+}}
{{ d^2\bDelta}}} = 
 \frac{ \alpha_S({{\bp^2}})}{ 2}
 f(x,{{ \bDelta}} )\nonumber\\
&\times& 
\Bigl\{\bigl|\Psi(z_g,\bp_g) -
\Psi(z_g,{{\bp_g}} -
{{\bDelta }})\bigr|^2 + \bigl|\Psi(z_g,\bp_{g}) -
\Psi(   z_g,\bp_{g}  +z\bDelta)\bigr|^2
{{\Bigr\}}} \,,
\label{eq:6.2}
\eea
The third component in (\ref{eq:6.1}) is the contribution from the coherent
diffractive excitation $q^*\to (qg)A$, which gives the exactly back-to-back
dijets. Here the nuclear attenuation factor is a consequence of the
initial parton $q^*$ having been colored one. The term in (\ref{eq:6.1}) 
is a counterpart of the second term in the free-nucleon spectrum, it 
satisfies the linear $k_{\perp}$-factorization in terms of $\phi(\bDelta)$
Finally, the first component of the free nucleon 
spectrum (\ref{eq:6.2})  gives rise to the nonlinear $k_{\perp}$-factorization
component in the nuclear spectrum  (\ref{eq:6.1}), which resembles strongly
the
truly inelastic dijet spectrum (\ref{eq:5.3}) for DIS. 
The principal difference is in the 
nuclear thickness dependence  of
the distortion factor: the asymmetric one, 
${{\Phi}}({{(2-\beta)}}\nu_A({{\bb}}),x_A,{{\bDelta}} -{{\bkappa_1}} -{{\bkappa}})
{{\Phi}}({{(1-\beta)}}\nu_A({{\bb}}),x_A,{{\bkappa_1}})$ for the fragmentation of colored
quark $q^*$ vs. the symmetric one, ${{\Phi}}({{(1-\beta)}}\nu_A({{\bb}}),x_A,{{\bDelta}} -{{\bkappa_1}} -{{\bkappa}})
{{\Phi}}({{(1-\beta)}}\nu_A({{\bb}}),x_A,{{\bkappa_1}})$ in DIS. In DIS it describes equal
distortion of the 
both outgoing parton waves by pure FSI, for the incident quarks $q^*$ in $pA$ 
collisions it includes the ISI distortion of the incoming wave of the colored quark $q^*$.
Subject to slight modifications for the color-representation dependence 
of the collective nuclear glue, the decomposition  (\ref{eq:6.1}) of the nuclear 
dijet spectrum will hold for dijet production via other pQCD subprocesses
like heavy flavour excitation $g^*\to Q\bar{Q}$ or gluon splitting $g^*\to gg$.


\section{The fate of $k_{\perp}$-factorization for nuclear targets: single-jet spectra
in $pA$ collisions}

The recovery of linear $k_{\perp}$-factorization (\ref{eq:3.5}) for
the single-jet spectrum in DIS is rather an exception due to the
abelianization in the case of a colorless projectile - the photon. 
The radiation of gluons from quarks, $q^*\to qg$, illustrates nicely
the salient features of breaking of  linear $k_{\perp}$-factorization
for the single-jet spectrum \cite{SingleJet}. It is directly relevant 
to jet production in the proton hemisphere of $pA$ collisions at RHIC
\cite{STAR,BRAHMS}.

Here we again show the large-$N_c$ results.
The spectrum of gluons for the free-nucleon target reads
\bea 
&& \frac{(2\pi)^2 d\sigma_A(q^* \to g(\bp_g) q) }{ dz_g d^2\bp_g }=
 \frac{1}{ 2}\int d^2\bkappa f(x,\bkappa) \nonumber\\
&&\Big\{ \bigl|
\Psi(z_g,\bp_g) - \Psi(z_g,\bp_g+\bkappa)\bigr|^2 + \bigl|\Psi(z_g,\bp_g+\bkappa) -
\Psi(z_g,\bp_g+z_g \bkappa)\bigr|^2 \Big\} \, .
\label{eq:7.1}
\eea
The same spectrum for the nuclear target is of the two-component form
\bea &&\frac{ (2\pi)^2 d\sigma_A(q^* \to g(\bp_g) q) }{ dz_g d^2\bp_g
d^2\bb} = \exp\left[-\nu_A(\bb)\right]\int d^2\bkappa \phi(\bb,x_A,\bkappa)
\nonumber\\
&&\Big\{|\Psi(z_g,\bp_g) - \Psi(z_g,\bp_g+\bkappa) |^2
+ |\Psi(z_g,\bp_g+\bkappa) - \Psi(z_g,\bp_g+z_g \bkappa)|^2\Big\} \\
&&+ \int d^2\bkappa_1 d^2\bkappa_2 \phi(\bb,x_A,\bkappa_1) \phi(\bb,x_A,\bkappa_2)
|\Psi(z_g,\bp_g+z_g \bkappa_1) - \Psi(z_g,\bp_g+\bkappa_1 + \bkappa_2)|^2 \, .
\nonumber\label{eq:7.2}
\eea
The first component is an exact counterpart of the free-nucleon spectrum 
subject to the familiar substitution $ f(x_A,\bkappa)\to  \phi(\bb,x_A,\bkappa)$.
It is suppressed by the nuclear absorption factor and for central interactions
the spectrum is entirely dominated by the second component which is a non-linear 
functional of the collective nuclear glue.
 
For soft gluons , $z_g \ll 1$,
the result (\ref{eq:7.2}) simplifies to
\bea
&&\frac{ (2\pi)^2 d\sigma_A(q^* \to g(\bp_g) q) }{ dz_g d^2\bp_g
d^2\bb}
= 
\int d^2\bkappa \,
\phi_{gg}(\bb,x_A,\bkappa)
|\Psi(z_g,\bp_g) - \Psi(z_g,\bp_g+\bkappa) |^2\, .
\label{eq:7.3}
\eea
In takes the linear $k_{\perp}$-factorization form in
terms of $\phi_{gg}(\bb,x_A,\bkappa)=\Big(\phi \otimes \phi\Big)(\bb,x_A,\bkappa)$
which has a meaning of the collective nuclear glue defined in terms of 
the intranuclear propagation of the gluon-gluon color dipole. This
illustrates nicely the important point that the collective nuclear glue is 
a density matrix in the color space rather than a single scalar function 
\cite{Nonlinear}.

One more point is noteworthy: there is a conspicuous difference
between the  $z_g$-dependence of the free-nucleon and nuclear spectra.
This amounts to the $\bp_g$-dependence of the Landau-Pomeranchuk-Migdal
effect; the same applies to the spectrum of leading quarks and
nuclear quenching of forward jets in $pA$ collisions \cite{SingleJet}. 

  
\section{Small-x evolution of collective nuclear glue.}

Despite the manifest breaking of the linear
$k_{\perp}$-factorization,
the collective nuclear glue remains a useful concept. 
For a free nucleon
target 
 the effect of the $q\bar{q}g$ and higher Fock states in the photon
{{is reabsorbed}} in 
the {{linear}} BFKL evolution for 
the dipole cross section with the photon 
treated as the $q\bar{q}$ state.
One possible definition of the nonlinear 
BFKL evolution for the nuclear unintegrated glue is to
insist on the same representation for the nuclear
profile function $\Gamma_A({{\bb} },x,{\br})$. 
It is indeed possible although
without a closed-form 
evolution equation. 

We comment here on the 
{{first iteration}} 
of the $\log\frac{1}{x}$ evolution. The modification of
the color dipole-nucleus profile function for the
$q\bar{q}g$ Fock state in the photon equals
\bea
\int d^2{{\bb} } 
\frac{\partial \delta \Gamma_A({{\bb} },x,{\br}) 
}{ \partial \log\frac{1}{ x}} =  
K_0 \int d^2{{\brho}} 
\frac{{{\br}}^2 }{ {{\brho}}^2
({{\brho}}-{{\br}})^2}
2\int d^2{{\bb} } 
[\Gamma_{3A}({{\bb}},{{\brho}},{{\br}}) 
-\Gamma_{2A}({{\bb}},{{\br}})]
\eea
\bea
 \Gamma_{3A}({{\bb} },{{\brho}},
{{\br}})&=& 1- 
S_{3A}({{\bb} },{{\brho}},
{{\br}}) 
=1-\exp[-\frac{1}{ 2}\sigma_1({{\brho}},{{\br}})
T({{\bb} })]
\eea
where $\sigma_3({{\brho}},{{\br}})$ is the 3-parton cross
section \cite{NZ94}.

A simplified Glauber-Gribov formula holds at large-$N_c$,
$
S_{3A}({{\bb} },{{\brho}},
{{\br}}) =S_{2A}({{\bb} },
{{\brho}}-{{\br}})
S_{2A}({{\bb} },{{\brho}}).
$
Here  ${\partial \Gamma_A({{\bb} },x,{\br})/
\partial \log\frac{1}{ x}}$ is a {{nonlinear}} 
functional of $\Gamma_{2A}$, the 
identification of $\Gamma_A(x,{{\bb} },{{\br}})$
with  $\Gamma_{2A}(x,{{\bb} },{\br})$, and
the 
extension of the {{first
iteration}} to the closed-form nonlinear equation  
as claimed in ref. \cite{BalitskiKovchegov} is  
{{unwarranted}}.
In terms of the nuclear transparency for large dipoles, $
S_{A}({{\bb} },\sigma_{0})=\exp[-\nu_A(\bb)], $ 
the {{first iteration}} for unintegrated nuclear glue
takes the form
\bea
\frac{\partial {\delta\phi_A({{\bb} },x,\bDelta)} 
}{ \partial \log\frac{1}{ x}} &=&S_{A}({{\bb} },\sigma_{0}) 
{{{\cal K}_{BFKL}}}\otimes 
{{\phi(\bb,x_A,\bDelta)}}\nonumber\\
&+&K_0\int d^2 \bp d^2\bk{{\phi}}(\bb,x_A,\bk)\nonumber\\
&\times& \Bigl\{K({{\bDelta}}+\bp,
{{\bDelta}}+\bk){{\phi}}(\bb,x_A,\bp)
-
K(\bp,\bp+{{\bDelta}}+\bk){{\phi(\bb,x_A,\bDelta)}}
\Bigr\}\nonumber\\
&=&S_{A}({{\bb} },\sigma_{0}) 
{{{\cal K}_{BFKL}}}\otimes 
{{\phi(\bb,x_A,\bDelta)}} +  
{{{\cal K}_{NonLin}}}\bigl[ 
{{\phi(\bb,x_A,\bDelta)}}\bigr]
\label{eq:BFKL1}
\eea
where
$K(\bp,\bk) = (\bp-\bk)^2/\bp^2\bk^2$. It contains an {{absorption 
suppressed linear BFKL  term}} with the familiar
kernel  ${{{\cal K}_{BFKL}}}$ \cite{BFKL}. 
For central DIS off heavy nuclei $S_{A}\to 0$ and
 evolution is entirely
driven by the nonlinear term quadratic in ${{\phi}}(\bb,x_A,\bk)$.

Making use of an explicit form of $K(\bp,\bk)$, one can recast 
(\ref{eq:BFKL1}) for the leading conformal twist nuclear glue in an alternative 
form 
\bea \frac{\partial \phi({{\bDelta}},x\bb) }{ \partial \log(1/x)} =
{\cal{K}}_{BFKL}\otimes \phi(\bb,x_A,{{\bDelta}}) +
{{{\cal{Q}}}}[\phi](\bb,x_A,{{\bDelta}}) \, . 
\label{eq:BFKL2}
\eea
Here the linear term evolves with the conventional BFKL kernel,
whereas the nonlinear term takes a particularly simple form
\bea {{{\cal{Q}}}}[\phi](\bb,x,{{\bDelta}}) = 
 &-&\frac{2 K_0 }{ {{\bDelta}}^2} \Big[ \int_{{{\bDelta^2}}} d^2 \bq
\phi(\bb,x_A,\bq) \Big]^2 \nonumber\\
&-& 2K_0 \phi(\bb,x_A,{{\bDelta}}) \int_{{{\bDelta}}^2} 
\frac{d^2\bp
}{ \bp^2} \int_{\bp^2} d^2\bq \phi(\bb,x_A,\bq)  .
\eea
For hard gluons, {{$\bDelta^2 > Q_A^2$}}, one can use an
approximation
$
\phi(\bq) \sim \phi({{\bDelta}}) \left({{{\bDelta}}^2 / q^2}\right)^2
$
with the result 
\bea {{{\cal{Q}}}}[\phi]({{\bDelta}};\bb) \approx -4 K_0 \cdot {{\bDelta}}^2
\phi^2 ({{\bDelta}})\propto 
\frac{\phi({{\bDelta}})}{{{\bDelta}}^2}\,.
\eea
The nonlinear component in (\ref{eq:BFKL2}) gives a pure higher
twist contribution. It doesn't exhaust the nuclear higher twist
terms, though, because the one is contained also in $\phi(x,{{\bDelta}};\bb)$,
see the discussion in \cite{Nonlinear,NSSdijet}.
The character of nonlinearity in terms of 
$ G_A({{\bb}},x_A,{{Q}})$ is instructive:
\bea
\frac{\partial^2{{\delta G_A}}({{\bb}},x,{{Q}})
}{
\partial\log(1/x)\partial \log{{Q^2}} } ={\cal{K}}_{BFKL}\otimes 
\frac{\partial{{G_A}}({{\bb}},x_A,{{Q}})
}{
\partial\log{{Q^2}} }
-\frac{4\alpha_S({{Q}}^2) T({{\bb}})
}{{{Q}}^2} \cdot \left(\frac{\partial{{G_A}}({{\bb}},x_A,{{Q}})
}{
\partial\log{{Q^2}} }\right)^2
\eea

Now have  a look at the plateau region of soft gluons , ${{\bDelta^2}} \ll Q_A^2$.
Here eq.~(\ref{eq:BFKL1}) takes the form
\bea
\frac{\partial {\phi_A({{\bb} },x,\bDelta)} 
}{ \partial \log\frac{1}{ x}} =-2C\pi K_0 
{{\phi({{\bb} },x_A,0)}}\label{eq:BFKLsoft}
\eea
where the constant factor, $C \sim 1$, depends on the form of the collective 
nuclear glue. If we recall that
\bea
{{\phi(x,{{\bb} },0)}} \sim \frac{1}{\pi 
{{Q_A^2}}({{\bb} },x)}
\eea
then (\ref{eq:BFKLsoft}) entails an
{{expansion of the plateau}} width with the decrease of $x$:
\bea
 {{Q_A^2}}({{\bb} }) 
\Longrightarrow  {{Q_A^2}}({{\bb} })
\left[ 1+2C\pi K_{0} \log\frac{1}{ x} \right] 
\eea
The full-fledged nonlinear evolution will be in effect for 
soft-to-hard intermediate gluon momenta ${{\bDelta^2}} \lsim Q_A^2$.

\section{CONCLUSIONS}
Nuclear saturation is a straightforward consequence of 
{{opacity of heavy nuclei}} to large color dipoles. 
The imposition of unitarity constraints within the color-dipole
approach leads to a unique definition and expansion of nuclear 
unintegrated glue in terms of {{\emph{the collective
glue of overlapping nucleons}}}. The problem of {{\emph{a non-abelian}}} 
intranuclear evolution of color dipoles has been solved and a
consistent description of single-jet and dijet production in DIS off nuclei
and hadron-nucleus collisions
 has been developed. We have proven the 
{{\emph{breaking $k_{\perp}$ factorization}}} and instead 
formulated the {{\emph{nonlinear $k_{\perp}$-factorization}}} for forward 
single-jet and dijet production. The {{\emph{nonlinear $k_{\perp}$-factorization}}} 
emerges as a universal feature of the pQCD description of
hard scattering in nuclear environment, still its mathematical 
formulation depends on the relevant pQCD subprocess. We applied our technique to
the nonlinear BFKL evolution of collective nuclear glue and 
explored the twist properties of the nonlinear component of this
equation.
\\

\noindent
{\bf Acknowledgements:} We are grateful to Lev N. Lipatov  for the 
invitation to, and a chance to present our ideas at, this exciting meeting.

\end{document}